\documentclass[a4paper,12pt]{article}

\usepackage{amsmath,amssymb}
\usepackage[dvipdfmx]{graphicx}
\numberwithin{equation}{section}
\usepackage{cite}
\usepackage{color}
\usepackage{ascmac}
\usepackage[breaklinks=true,linktocpage=true]{hyperref}
\usepackage{mathrsfs}
\usepackage[top=2.5cm,bottom=2.5cm, left=2.5cm,right=2.5cm]{geometry}
\usepackage{theorem}

\theorembodyfont{\upshape}

\newcommand{\Ga}{{\Gamma}}
\newcommand{\De}{{\Delta}}

\newcommand{\Sig}{{\Sigma}}

\newcommand{\al}{{\alpha}}
\newcommand{\bt}{{\beta}}
\newcommand{\ga}{{\gamma}}
\newcommand{\de}{{\delta}}

\newcommand{\lm}{{\lambda}}

\newcommand{\sig}{{\sigma}}
\newcommand{\vphi}{{\varphi}}

\newcommand{\bbD}{{\mathbb{D}}}

\newcommand{\cD}{{\mathcal{D}}}
\newcommand{\cM}{{\mathcal{M}}}

\newcommand{\dg}{{\dagger}}

\newcommand{\lan}{{\langle}}

\newcommand{\nn}{{\nonumber}}

\newcommand{\pd}{{\partial}}

\newcommand{\ran}{{\rangle}}

\newcommand{\wt}{\widetilde}

\def\bbra{{\langle\kern-2.5pt\langle}}
\def\kket{{\rangle\kern-2.5pt\rangle}}
\def\Bbra{{\Big\langle\kern-3.5pt\Big\langle}}
\def\Kket{{\Big\rangle\kern-3.5pt\Big\rangle}}

\makeatletter
\@addtoreset{equation}{section}
\makeatother

\begin{document}

\thispagestyle{empty}

\begin{center}
{\Large{\textbf{On conformal correlators and blocks with spinors \\ in general dimensions}}}
\\
\medskip
\vspace{1cm}
\textbf{
Hiroshi~Isono 
}
\bigskip

{\small Department of Physics, Faculty of Science, Chulalongkorn University,
\\Phayathai Road, Pathumwan, Bangkok 10330, Thailand}

\bigskip

{\small e-mail: \texttt{hiroshi.isono81@gmail.com}}

\end{center}

\vspace{1cm}

\begin{abstract}
We compute conformal correlation functions with spinor, tensor, and spinor-tensor primary fields 
in general dimensions with Euclidean and Lorentzian metrics. 
The spinors are taken to be Dirac spinors, which exist for any dimensions.
For this, the embedding space formalism is employed
and the polarisation spinors are introduced to simplify the computations.
Three-point functions are rewritten in terms of differential operators 
acting on scalar-scalar-tensor correlation functions.
This enables us to determine conformal blocks for four-point functions with scalar and spinor fields
by acting the differential operators on scalar conformal blocks,
which will be useful in finding their geodesic Witten diagrams.
\end{abstract}

\newpage

\tableofcontents

\section{Introduction}
Rapid development of conformal field theories in recent years
is based on several characterisations of conformal blocks, including 
double series expansion forms by the direct summation over descendant fields
\cite{Ferrara:1971vh, Ferrara:1973yt, Ferrara:1973vz,Ferrara:1974nf,Ferrara:1974ny,Petkou:1994ad,Dolan:2000ut},
solutions to Casimir differential equations \cite{Dolan:2003hv,Dolan:2011dv},
Mellin space representations \cite{Mack:2009mi,Fitzpatrick:2012cg,Gopakumar:2016cpb},
conformal integrals with shadow fields followed by monodromy projections \cite{SimmonsDuffin:2012uy}.

In addition, recent discovery of the geodesic Witten diagrams for conformal blocks
\cite{Hijano:2015zsa}
is remarkable from the viewpoint of AdS/CFT correspondence.\footnote{
The author is also interested in the possibility of its de Sitter space version,
from the viewpoint of applications to primordial correlation functions in inflationary cosmology;
in particular, decompositions of primordial correlators along the line of \cite{Arkani-Hamed:2015bza}, 
holographic computations of non-Gaussianities 
using three-dimensional conformal perturbation theory like 
\cite{Schalm:2012pi,Bzowski:2012ih,Garriga:2013rpa,Garriga:2014ema,Isono:2016yyj}. 
}
A geodesic Witten diagram for a conformal block at $P_1,...,P_4$ 
for a $d$-dimensional conformal field theory (CFT$_d$)
is a tree-level exchange Witten diagram in the $(d+1)$-dimensional anti-de Sitter space (AdS$_{d+1}$),
in which two bulk interaction vertices are integrated over two geodesics, 
one of them terminating at $P_1,P_2$ in the AdS$_{d+1}$ boundary
and the other at $P_3,P_4$.
So far, geodesic Witten diagrams for conformal blocks 
associated with four-point functions with tensors of arbitrary integer spins in general dimensions
have been studied
\cite{Hijano:2015zsa,Nishida:2016vds,Castro:2017hpx,Dyer:2017zef,Sleight:2017fpc,Chen:2017yia,Rastelli:2017ecj}.
One way to derive the geodesic Witten diagram for a conformal block
\cite{Castro:2017hpx,Dyer:2017zef,Sleight:2017fpc,Chen:2017yia}
employs the shadow field formalism 
\cite{Ferrara:1972xe,Ferrara:1972ay,Ferrara:1972uq,Ferrara:1973vz,SimmonsDuffin:2012uy},
in which a conformal block for a primary field $O$ 
is written as the integral of the product of two three-point functions,
one with $O$ and the other with its shadow field,
followed by a monodromy projection to extract the correct conformal block.
In this derivation, \cite{Castro:2017hpx,Dyer:2017zef,Sleight:2017fpc,Chen:2017yia}
used differential representations of three-point functions with arbitrary integer spins,
in which differential operators act on scalar three-point correlation functions,
so that any other conformal blocks are obtained by acting appropriate differential operators 
on the geodesic Witten diagrams for scalar conformal blocks.
This proof makes thorough use of the embedding space formalism for tensor fields \cite{Costa:2011mg}, 
which simplifies the computations of correlation functions with arbitrary spins drastically.

A chief motivation of our work is to find geodesic Witten diagrams for conformal blocks 
associated with four-point functions involving spinor fields in general dimensions.
As the first step in this direction, this paper discusses three-point functions 
with spinor fields (i.e. spin 1/2), tensor primary fields of arbitrary integer spins,
and spinor-tensor primary fields of arbitrary half-integer spins.
They have been studied in some specific dimensions
under some constraints on spinors such as chirality, Majorana conditions, 
in, for example, \cite{Petkou:1996np,Weinberg:2010fx,Iliesiu:2015qra,Iliesiu:2015akf,Cuomo:2017wme},\footnote{
After submission to arXiv, we noticed that the recent paper \cite{Cuomo:2017wme}
gives detailed analysis of four-dimensional correlation functions with spinors, 
where spinors are represented with the standard dotted and undotted indices,
which are peculiar to four dimensions.
}
and, in the context of AdS/CFT correspondence, in, for example,
\cite{Henningson:1998cd,Mueck:1998iz,Ghezelbash:1998pf,Corley:1998qg,Volovich:1998tj,Koshelev:1998tu,Kawano:1999au}.
In this paper, we compute three-point functions 
in general dimensions, without imposing any conditions on spinors,
so that we work with Dirac spinors, which exist in any dimensions.
In general, three-point functions with spinors 
are linear combinations of more than one contributions with free parameters, 
which are called three-point structures.
We first find all three-point structures for three-point functions with spinors,
and then write them in differential representations,
keeping in mind their applications to geodesic Witten diagrams.

This paper is organised as follows.
In Section \ref{spinor}, we give a brief review of the embedding formalism, 
with particular emphasis on properties of spinor primary fields in the embedding formalism.
In Section \ref{correlators}, we compute the following three-point functions.
One is for two spinor fields and one tensor field of arbitrary integer spin, 
and the other is for one spinor field, one scalar field and one spinor-tensor field of arbitrary half-integer spin.
In contrast to \cite{Iliesiu:2015qra,Iliesiu:2015akf,Cuomo:2017wme}, 
we keep as much vector indices as possible
in tensor fields and spinor-tensor fields, to avoid a laborious task of converting spinor indices to tensor ones.
We then find their differential representations,
and, based on them, derive general formulae for conformal blocks associated with four-point functions 
with spinor and scalar fields.
Section \ref{conclusion} gives the conclusion and future directions, 
with emphasis on the derivation of geodesic Witten diagrams with AdS spinor propagators.
Appendix \ref{app:pol-tensor} gives a brief summary of tensor fields and their correlation functions
in the embedding space formalism.
Appendix \ref{app:embed} shows our conventions for the Dirac matrices.
Appendix \ref{app:antisymmetric} presents an example of three-point functions 
of two spinor and one tensor field with antisymmetric indices.

\section{Spinors in conformal field theories}
\label{spinor}
In this section, we introduce spinor fields in conformal field theories in general dimensions, 
basically following \cite{Weinberg:2010fx,Iliesiu:2015qra,Iliesiu:2015akf}. 
The embedding space formalism is employed to simplify our analysis.

\subsection{Embedding space}
The embedding space provides a natural framework for conformal field theories 
because the conformal group is realised linearly in the embedding space.
This section first provides a brief review of the embedding space formalism. 

Let $M$ be a $d$-dimensional Minkowski spacetime with a flat metric $\eta_{\mu\nu}$
in which a $d$-dimensional conformal field theory lives, 
which we call the physical space.\footnote{
Indices $\mu,\nu,...=1,2,...,d$ refer to the physical space,
while indices $M,N,...=1,2,...,d,\sharp,\flat$ refer to the embedding space.
}
The embedding space $\Sig$ is a $(d+2)$-dimensional flat spacetime with a flat metric $\eta_{MN}$.
If $\eta_{\mu\nu}$ has $s$ plus and $t$ minus signs, 
then $\eta_{MN}$ is defined to have $(s+1)$ plus and $(t+1)$ minus signs.
More explicitly, the two metrics are common for $(\mu\nu)$ components,
and the new components of $\eta_{MN}$ are given by
\begin{align}
\eta_{\sharp\sharp}=1, \quad \eta_{\flat\flat}=-1, \quad \eta_{\sharp\flat}=0.
\end{align}
The physical space $M$ is embedded into the embedding space $\Sig$ 
as the null hypersurface
\begin{align}
P^2:=\eta_{MN}P^MP^N=0.
\end{align}
It is convenient to introduce the light-cone coordinates $P^\pm=P^\flat \pm P^\sharp$.
In what follows, we will let $P^M$ represent the light-cone coordinate $(P^\mu,P^+,P^-)$.
In terms of the light-cone coordinates, the null hypersurface equation reads
\begin{align}
P^\mu P_\mu = P^+P^-.
\end{align}
The physical space $M$ is isomorphic to the projection of the null hypersurface 
by the equivalence relation $P \sim \lm P$,
meaning that a point $x^\mu$ in $M$ corresponds to a line $\lm(x^\mu, 1, x^2)$ in the null hypersurface.
We will denote representative $(x^\mu,1,x^2)$ by $P^M(x)$.
Conversely, a given point $P^M$ in the null hypersurface corresponds to $P^\mu/P^+$ in $M$.
This establishes the isomorphism between $M$ and the null hypersurface.

\subsection{Conformal primary fields in embedding space formalism}
\label{primary}
In this subsection, we introduce conformal primary fields in the physical space
and embed them into the embedding space.
For this purpose, it is useful to review the conformal algebra and then define primary fields, 
following \cite{Weinberg:2010fx}. 
The $d$-dimensional conformal algebra can be reorganised into the $(d+2)$-dimensional Lorentz algebra,
\begin{align}
i[M^{PQ}, M^{RS}] = \eta^{QR}M^{PS} + \eta^{PS}M^{QR} - \eta^{PR}M^{QS} - \eta^{QS}M^{PR}.
\end{align}
The generators of the $d$-dimensional conformal algebra, defined by
\begin{align}
M^{\mu\nu}, \quad
P^\mu := M^{\flat\mu}+M^{\sharp\mu}, \quad
K^\mu := M^{\flat\mu}-M^{\sharp\mu}, \quad
D := M^{\flat\sharp},
\end{align}
satisfy the following conformal algebra\footnote{
Note that the translation generator $P^\mu$ has nothing to do with the embedding coordinate $P^M$.
This will not be confusing because the translation generator will not be used except in this subsection.
}
\begin{align}
i[M^{\mu\nu}, M^{\rho\sig}] 
&= \eta^{\nu\rho}M^{\mu\sig}+\eta^{\mu\sig}M^{\nu\rho}-\eta^{\mu\rho}M^{\nu\sig}-\eta^{\nu\sig}M^{\mu\rho}, \\
i[M^{\mu\nu}, P^\rho] &= \eta^{\nu\rho}P^\mu - \eta^{\mu\rho}P^\nu, \\ 
i[M^{\mu\nu}, K^\rho] &= \eta^{\nu\rho}K^\mu - \eta^{\mu\rho}K^\nu, \\ 
i[D,P^\mu] &= P^\mu, \\
i[D,K^\mu] &= -K^\mu, \\
i[K^\mu, P^\nu] &= 2(M^{\mu\nu}-\eta^{\mu\nu}D).
\end{align}
A general primary field $\vphi_n$ is characterised by the following commutation relations with the conformal generators,
\begin{align}
i[M^{\mu\nu}, \vphi_n(x)] &= (x^\nu\pd^\mu-x^\mu\pd^\nu)\vphi_n(x) - i(\cM^{\mu\nu})_{nm}\vphi_m(x), 
\label{Mphi} \\
i[P^\mu, \vphi_n(x)] &= -\pd^\mu\vphi_n(x), \label{Pphi} \\ 
i[K^\mu, \vphi_n(x)] &= 
(2x^\mu x^\nu \pd_\nu-x^2 \pd^\mu + 2\De x^\mu) \vphi_n(x)
-2i x_\nu (\cM^{\nu\mu})_{nm}\vphi_m(x), 
\label{Kphi} \\
i[D, \vphi_n(x)] &= (x^\mu\pd_\mu+\De)\vphi_n(x),
\label{Dphi}
\end{align}
where the matrices $(\cM^{\mu\nu})_{mn}$ give a representation of the Lorentz generators $M^{\mu\nu}$.
The matrices for scalar fields are given by $\cM^{\mu\nu}=0$.
For vector fields, we can use 
$i(\cM^{\mu\nu})^\rho{}_\sig=\eta^{\mu\rho}\de^\nu{}_\sig-\eta^{\nu\rho}\de^\mu{}_\sig$,
and the extension to more general tensor fields is straightforward.
The spinor representation is given by
\begin{align}
\cM^{\mu\nu} = -\frac{i}{2}\ga^{\mu\nu} = -\frac{i}{4}[\ga^\mu, \ga^\nu],
\end{align}
where we introduced the $d$-dimensional Dirac matrices satisfying
\begin{align}
\{ \ga_\mu, \ga_\nu \} = 2\eta_{\mu\nu}.
\end{align}

This paper will deal with correlation functions with tensor fields, spinor fields, and spinor-tensor fields.
Correlation functions of tensor primary fields have long been studied extensively.
Especially, computations of them in the embedding formalism \cite{Costa:2011mg} are of special relevance to us.
We will not give their detailed properties but just give a brief summary in Appendix \ref{app:pol-tensor}.
In the rest of this section, we explain spinor and spinor-tensor primary fields in the embedding space formalism.

\subsubsection{Spinor primary fields}
We proceed to embedding spinor primary fields in the embedding space.
As explained above, the physical space $M$ is identified 
with the projection of the null hypersurface in the embedding space $\Sig$,
in which two points $P$ and $\lm P$ are identified.
It is therefore reasonable to relate the value of a field at $P$ with that at $\lm P$. 
In the embedding space formalism, 
this relation determines the conformal dimension of the corresponding primary field in $M$. 
We will call this relation the homogeneity condition.

To embed a spinor primary field of spin $1/2$ into the embedding space,
we first introduce a spinor field $\Psi(P)$ in the embedding space,
which is characterised by the following commutation relations with the Lorentz generators
\begin{align}
\label{comPsi}
i[M^{MN}, \Psi(P)] = (P^N\pd^M-P^M\pd^N)\Psi(P) - \frac{1}{2}\Ga^{MN}\Psi(P),
\end{align}
where we introduced the $(d+2)$-dimensional Dirac matrices $\Ga^M$ 
satisfying $\{ \Ga_M, \Ga_N \} = 2\eta_{MN}$.
Since our target is the $d$-dimensional CFT, 
it is convenient to express the $(d+2)$-dimensional Dirac matrices using the $d$-dimensional ones.
This depends on the choice of $\eta_{\mu\nu}$. 
In this paper we consider the following two typical cases: 
$\eta_{\mu\nu}$ with no minus sign, and with one $-1$.
Explicit forms of the $(d+2)$-dimensional Dirac matrices
are given in Appendix \ref{app:embed}, 
where explicit definitions of the Dirac conjugate matrices are also given.
In what follows, we work in the case with the Lorentzian metric.
The Euclidean case will be explained later. 

Let us now proceed to a prescription for recovering the original primary field $\psi(x)$ 
from the embedding space spinor $\Psi(P)$, based on \cite{Iliesiu:2015qra,Iliesiu:2015akf}.
We first impose on $\Psi$ the following homogeneity condition
\begin{align}
\label{homoPsi}
\Psi(\lm P) = \lm^{-\De-\frac{1}{2}} \Psi(P).
\end{align}
A point $x$ in $M$ is identified with a $(P^\mu,P^+,P^-)$ in $\Sig$ under the relation $x^\mu=P^\mu/P^+$.
Based on this identification, let us construct a field $\psi_\pm$ in $M$ by
\begin{align}
\psi_\pm(x) := (P^+)^{\De+\frac{1}{2}}\Psi_\pm(P) =  \Psi_\pm(P(x)),
\end{align}
where the spinor $\Psi$ is decomposed as $\Psi=(\Psi_+,\Psi_-)^T$, where $\Psi_+$ and $\Psi_-$ have the same size.
Using the commutation relation \eqref{comPsi} for the embedding space spinor at $P^M=P^M(x)$,
we can compute the commutation relations of $\psi_\pm$ with the conformal generators,\footnote{
In the derivations, we used the following identities to express differential operators on $\Psi_\pm$ 
in those on $\psi_\pm$,
\begin{align}
&{}
\pd_\mu\psi_\pm(x)
= \pd_\mu \Psi_\pm(x^\mu,1,x^2)
= (\pd_\mu + 2x_\mu\pd_-) \Psi_\pm(P)|_{P=P(x)}, \nn\\
&{}
\left(-\De-\tfrac{1}{2}\right)\psi_\pm(x) = \left( x^\mu\pd_\mu + \pd_+ + x^2\pd_- \right)\Psi(P)|_{P=P(x)}, \nn
\end{align}
where the differential operators acting on $\psi_\pm$ are with respect to $x^\mu$, 
while those acting on $\Psi_\pm$ are with respect to $P^M$.
Note that the second identity comes from the homogeneity condition \eqref{homoPsi}.
}
\begin{align}
i[M^{\mu\nu}, \psi_+(x)] &= 
(x^\nu\pd^\mu-x^\mu\pd^\nu)\psi_+(x)-\tfrac{1}{2}\ga^{\mu\nu}\psi_+(x), \\
i[M^{\mu\nu}, \psi_-(x)] &= 
(x^\nu\pd^\mu-x^\mu\pd^\nu)\psi_-(x)-\tfrac{1}{2}\de\ga^{\mu\nu}\de^{-1}\psi_-(x), \\
i[P^\mu, \psi_+(x)] &= -\pd^\mu\psi_+(x) + \ga^\mu \de^{-1}\psi_-(x), \\
i[P^\mu, \psi_-(x)] &= -\pd^\mu\psi_-(x), \\
i[K^\mu, \psi_+(x)] &= 
(2x^\mu x^\nu \pd_\nu - x^2\pd^\mu)\psi_+(x) + (2\De+1) x^\mu \psi_+(x), \\
i[K^\mu, \psi_-(x)] &= 
(2x^\mu x^\nu \pd_\nu - x^2\pd^\mu)\psi_-(x) + (2\De+1) x^\mu \psi_-(x) + \de\ga^\mu\psi_+(x), \\
i[D, \psi_\pm(x)] &= 
x^\mu\pd_\mu\psi_\pm(x) + (\De+\tfrac{1}{2})\psi_\pm(x) \mp \tfrac{1}{2}\psi_\pm(x),
\end{align}
Note that they are different from the commutation relations \eqref{Mphi}-\eqref{Dphi} 
for $d$-dimensional spinor primary fields.
It, however, turns out that the linear combination
\begin{align}
\psi_+-x^\rho\ga_\rho\de^{-1}\psi_-
\end{align}
satisfies the defining commutation relations \eqref{Mphi}-\eqref{Dphi} for a $d$-dimensional spinor primary field of dimension $\De$.

This result can be rephrased compactly by introducing the polarisation spinors \cite{Iliesiu:2015qra,Iliesiu:2015akf}. 
Let us first introduce a Grassmann-even spinor $s$ in the physical space,
to form a scalar field $\psi(x,s)=\bar s\psi(x)$ from a spinor primary field $\psi$ of dimension $\De$.
Here the Dirac conjugate of $s$ is defined by $\bar s=s^\dg\de$, 
where $\de$ is the Dirac conjugation matrix for $d$-dimensional spinors (see Appendix \ref{app:embed}).
Let us further introduce a polarisation spinor $S$ in the embedding space so that
\begin{align}
\bar S = \left( \bar s, -\bar s x^\rho\ga_\rho \de^{-1} \right).
\end{align}
Here the Dirac conjugate of $S$ is defined by $\bar S=S^\dg D$,
where $\de$ is the Dirac conjugation matrix for $(d+2)$-dimensional spinors (see Appendix \ref{app:embed}).
Using the explicit form of $D$, we find
\begin{align}
S = \begin{pmatrix} x^\rho\ga_\rho s \\ \de s \end{pmatrix}.
\end{align}
In terms of this, the identification of $\psi$ with $\psi_+ - x^\rho\ga_\rho\de^{-1}\psi_-$ can be rephrased as
\begin{align}
\psi(x,s) = \Psi(P(x),S) = (P^+)^{\De+\frac{1}{2}} \Psi(P,S),
\label{recover-spinor}
\end{align}
under the identification $x^\mu=P^\mu/P^+$. Here we defined a scalar field $\Psi(P,S):=\bar S\Psi(P)$.
This relation \eqref{recover-spinor} can be used to recover the original spinor $\psi(x)$ 
from the contracted one $\Psi(P,S)$ in the embedding space.

We here derive two important properties, 
which will play crucial roles in constructing conformal correlation functions.
First, the homogeneity condition \eqref{homoPsi} yields the following homogeneity condition on $\Psi(P,S)$
\begin{align}
\label{homoPS}
\Psi(aP, bS) = a^{-\De-\frac{1}{2}}b\Psi(P, S).
\end{align}
Second, the polarization spinor $S$ also satisfies the transversality condition
\begin{align}
\label{transvS}
\bar S P^M \Ga_M = 0.
\end{align}
The homogeneity condition \eqref{homoPS} and the transversality condition \eqref{transvS}, 
combined with the $(d+2)$-dimensional Lorentz invariance, 
provide strong constraints in the construction of correlation functions with spinor primary fields.

\subsubsection{Spinor-tensor primary fields}
Let us consider a primary field $\psi_{\mu_1...\mu_\ell}$ of spin $\ell+\frac{1}{2}$ on dimension $\De$, 
which has one spinor index and $\ell$ vector indices. 
To this, we assign an embedding-space spinor-tensor field $\Psi_{M_1...M_\ell}$ 
with one spinor index (not written explicitly) and $\ell$ vector indices, which are also symmetric.
As in the spin $1/2$ case, 
it is convenient to use the polarization spinor $S$, which was introduced above, to form a tensor field 
$\Psi_{M_1...M_\ell}(P,S)=\bar S\Psi_{M_1...M_\ell}(P)$.
The homogeneity condition imposed on this tensor field reads
\begin{align}
\label{homoPS-st}
\Psi_{M_1...M_\ell}(aP, bS) = a^{-\De-\frac{1}{2}}b\Psi_{M_1...M_\ell}(P, S),
\end{align}
and the polarisation spinor satisfies the transversality condition \eqref{transvS}.
Furthermore, we have one more constraint, 
which comes from transversality condition on tensor fields \eqref{transvT},
\begin{align}
\label{transv-st}
P^{M_k}\Psi_{M_1...M_\ell}(P,S)=0 \quad\quad \mbox{ for } \quad 1 \leq k \leq \ell.
\end{align}

\section{Correlation functions with spinor primary fields}
\label{correlators}
In this section, we compute correlation functions with spinor, symmetric traceless tensor 
and symmetric spinor-tensor primary fields.
In the embedding space, 
the correlation functions involve symmetric tensor field of spin $\ell$
and symmetric spinor-tensor field of spin $\ell+(1/2)$,\footnote{
As mentioned briefly in the Introduction, in contrast with \cite{Iliesiu:2015qra,Iliesiu:2015akf},
where tensor fields of integer spins $\ell$, half-integer spins $\ell+(1/2)$
are written as multispinor fields with $2\ell$ spinor indices and at most one vector index, 
we keep the vector indices as much as possible, as given in \eqref{embed-fields}.
}
\begin{align}
\label{embed-fields}
O_{M_1...M_\ell}(P), \quad\quad
O_{M_1...M_\ell}(P,S),
\end{align} 
where $\ell \geq 0$ is an integer.
Since the spinor indices are contracted with the polarization spinors, 
it is convenient to introduce Lorentz tensors containing two polarisation spinors 
\cite{Iliesiu:2015qra,Iliesiu:2015akf}, 
which we will call spinor quadratics. 
More explicitly, we denote a spinor quadratics with polarisation spinors $S_1,S_2$ by
\begin{align}
\lan \bar S_1 \,...\, S_2 \ran,
\end{align}
where $S_i$ is assigned to a spinor field $\Psi(P_i)$.
The inside of the bracket consists of the $(d+2)$-dimensional Dirac matrices
and/or vectors contracted with the Dirac matrices. 
For example, $\lan \bar S_1 \Ga_MP S_2 \ran$ means $\bar S_1 \Ga_M\Ga_NP^N S_2$.
In terms of spinor quadratics, 
the transversality condition \eqref{transvS} and its Hermitian conjugate become
\begin{align}
\lan \bar S_1P_1...S_2 \ran=0, \quad\quad \lan \bar S_1...P_2S_2 \ran=0.
\end{align}

Furthermore, to simplify computations of correlation functions, 
it is convenient to introduce the polarisation vector $Z_M$ to form scalar fields \cite{Costa:2011mg}; 
for example,
\begin{align}
O^{(\ell)}(P, Z) &= Z^{M_1}...Z^{M_\ell}O_{M_1...M_\ell}(P), \\
O^{(\ell+\frac{1}{2})}(P, S, Z) &= Z^{M_1}...Z^{M_\ell}O_{M_1...M_\ell}(S,P).
\end{align}
The original tensor fields are recovered from the contracted scalar fields 
by applying the differential operator \eqref{recover-tensor-diffop} to the scalar fields.
A brief summary of the polarisation vectors needed in this paper is given in Appendix \ref{app:pol-tensor}.\footnote{
For more details, see, for example, \cite{Costa:2011mg}.
}
Therefore, the strategy of our computations of correlation functions 
is first to compute scalar correlation functions with polarisation vectors $Z$ and spinors $S$,
and to recover the original correlation functions with explicit tensor and spinor indices 
using \eqref{recover-tensor} and \eqref{recover-spinor}.

\subsection{Two-point functions}
Let us first compute two-point functions of spinor or spinor-tensor fields,\footnote{
$\bar O(P,S)$ is defined by $\bar O(P,S)=\bar O(P) S$.
}
\begin{align}
\lan O^{(\ell+\frac{1}{2})}(P_1, S_1, Z_1) \bar O^{(\ell+\frac{1}{2})}(P_2, S_2, Z_2) \ran,
\end{align}
where $\ell \geq 0$, and the two fields have dimension $\De$.
We first consider odd-dimensional conformal field theories.
Since the fields in the correlation function are made scalar with the polarisation spinors and vectors,
it is a function of Lorentz scalars $P_{ij}:=-2P_i.P_j=x_{ij}^2$
and spinor quadratics with $\bar S_1,S_2$ which are Lorentz scalar.
Let us parametrise the correlation function as
\begin{align}
\label{fl+1/2l+1/2}
\lan O^{(\ell+\frac{1}{2})}(P_1, S_1, Z_1) \bar O^{(\ell+\frac{1}{2})}(P_2, S_2, Z_2) \ran
= f_{\ell+\frac{1}{2},\ell+\frac{1}{2}}(P_i,\bar S_1,S_2,Z_1,Z_2)
P_{12}^{-\De-\frac{1}{2}}.
\end{align}
The homogeneity conditions \eqref{homoPS-st}
and the transversality condition \eqref{transvTZ}
amount to the following constraints on $f_{\ell+\frac{1}{2},\ell+\frac{1}{2}}$
\begin{align}
&{}
f_{\ell+\frac{1}{2},\ell+\frac{1}{2}}(\lm_iP_i, a\bar S_1, bS_2, cZ_1, dZ_2) 
= abc^\ell d^\ell f_{\ell+\frac{1}{2},\ell+\frac{1}{2}}(P_i, \bar S_1, S_2,Z_1,Z_2), 
\label{homol+1/2l+1/2} \\
&{}
f_{\ell+\frac{1}{2},\ell+\frac{1}{2}}(\lm_iP_i, \bar S_1, S_2, Z_1+\al P_1, Z_2+\bt P_2) 
= f_{\ell+\frac{1}{2},\ell+\frac{1}{2}}(P_i, \bar S_1, S_2, Z_1, Z_2).
\label{transvl+1/2l+1/2}
\end{align}
The first condition implies that the function $f_{\ell+\frac{1}{2},\ell+\frac{1}{2}}$ 
must have only one spinor quadratics with $\bar S_1,S_2$. 
Taking into account the transversality condition on the polarisation spinors $\bar S_1,S_2$, 
we find that the solution to the conditions \eqref{homol+1/2l+1/2} and \eqref{transvl+1/2l+1/2} 
is uniquely (up to a numerical constant)
\begin{align}
\label{2pt1}
\lan \bar S_1S_2 \ran \left[ \frac{(Z_1.Z_2)P_{12} + 2(Z_1.P_2)(Z_2.P_1)}{P_{12}} \right]^\ell.
\end{align}
Let us move on to the even-dimensional case.
Since the even-dimensional Clifford algebra allows the chirality matrix, 
the conditions \eqref{homol+1/2l+1/2} and \eqref{transvl+1/2l+1/2} yield one more solution
\begin{align}
\label{2pt2}
\lan \bar S_1 \Ga S_2 \ran \left[ \frac{(Z_1.Z_2)P_{12} + 2(Z_1.P_2)(Z_2.P_1)}{P_{12}} \right]^\ell,
\end{align}
where $\Ga$ is the chirality matrix for $(d+2)$-dimensional Dirac matrices.
Therefore, the coefficient function $f_{\ell+\frac{1}{2},\ell+\frac{1}{2}}$ is a linear combination 
of \eqref{2pt1} and \eqref{2pt2}.
The appearance of the chirality matrix is discussed in \cite{Weinberg:2010fx} in terms of the parity.

\subsection{Three-point functions of two spinors and one tensor} 
\paragraph{Two spinors and one scalar}
Let us first consider a correlation function of two spinor and one scalar primary fields\footnote{
Here $(P_i)_M$ in the null hypersurface corresponds to $(x_i)_\mu$ in the physical space 
so that $P_M(x_i)=(P_i)_M$.
}
\begin{align}
\lan \Psi_1(P_1,S_1) \bar\Psi_2(P_2,S_2) O(P_3) \ran,
\end{align}
where $\Psi_{1(2)}$ has dimension $\De_{1(2)}$ and $O$ has dimension $\De_3$.
We first consider odd-dimensional conformal field theories.
Since the fields in the correlation function are made scalar with the polarisation spinors,
it is a function of Lorentz scalars $P_{ij}:=-2P_i.P_j=x_{ij}^2$
and spinor quadratics with $\bar S_1,S_2$ which are Lorentz scalar.
It is convenient to parametrise the correlation function as\footnote{
We will use $\Sig_{ij}=\De_i+\De_j$ and $\De_{ij}=\De_i-\De_j$.}
\begin{align}
\lan \Psi_1(P_1,S_1) \bar\Psi_2(P_2,S_2) O(P_3) \ran
= f_{\frac{1}{2}\frac{1}{2}0}(P_i,\bar S_1,S_2)
P_{12}^{-\frac{\Sig_{12}-\De_3+1}{2}}
P_{23}^{-\frac{\De_3-\De_{12}}{2}}
P_{31}^{-\frac{\De_3+\De_{12}}{2}}.
\end{align}
The homogeneity conditions on the three fields \eqref{homoPS}, \eqref{homoT}
amount to the following constraint on $f_{\frac{1}{2}\frac{1}{2}0}$,
\begin{align}
\label{f1/21/20}
f_{\frac{1}{2}\frac{1}{2}0}(\lm_iP_i, a\bar S_1, bS_2) 
= ab f_{\frac{1}{2}\frac{1}{2}0}(P_i, \bar S_1, S_2).
\end{align}
This condition implies that the function $f_{\frac{1}{2}\frac{1}{2}0}$ 
must have only one spinor quadratics with $\bar S_1,S_2$. 
Taking into account the transversality condition on the polarisation spinors $\bar S_1,S_2$, 
we find that solutions to the homogeneity condition \eqref{f1/21/20} 
are given as linear combinations of the following two quantities,
which we will call the three-point structures,\footnote{
Note first that $P_i^2=0$ inside spinor quadratics.
Spinor quadratics with more embedding coordinates $P_i$ are not included
because they become linear combinations 
of $f^{(1)}_{\frac{1}{2}\frac{1}{2}0}$ and $f^{(2)}_{\frac{1}{2}\frac{1}{2}0}$.
For the same reason, changing the ordering of $P_i$ inside the spinor quadratics 
adds nothing to \eqref{001/2odd} and \eqref{001/2even}.}
\begin{align}
\label{001/2odd}
f^{(1)}_{\frac{1}{2}\frac{1}{2}0}:=\lan \bar S_1S_2\ran, \quad\quad
f^{(2)}_{\frac{1}{2}\frac{1}{2}0}:=\lan \bar S_1P_3S_2 \ran \sqrt{\frac{P_{12}}{P_{13}P_{23}}}.
\end{align}
For even-dimensional conformal field theories, the condition \eqref{f1/21/20} yields two more solutions
\begin{align}
\label{001/2even}
f^{(3)}_{\frac{1}{2}\frac{1}{2}0}:=\lan \bar S_1 \Ga S_2\ran, \quad\quad
f^{(4)}_{\frac{1}{2}\frac{1}{2}0}:=\lan \bar S_1 \Ga P_3S_2 \ran \sqrt{\frac{P_{12}}{P_{13}P_{23}}},
\end{align}
The three-point structure $f^{(1)}$ is found in e.g. \cite{Iliesiu:2015qra},
while $f^{(2)}$ is found in e.g. \cite{Petkou:1996np}.

\paragraph{Two spinors and one tensor of spin $\ell$}
Let us consider a three-point correlation function 
with two spinors and one symmetric tensor field,
\begin{align}
\lan \Psi_1(P_1,S_1) \bar\Psi_2(P_2,S_2) O^{(\ell)}(P_3,Z_3) \ran,
\end{align}
where $\Psi_{1(2)}$ has dimension $\De_{1(2)}$ and $O$ has dimension $\De_3$ and integer spin $\ell$.
As in the previous case, we first consider odd-dimensional conformal field theories.
It is convenient to parametrise the correlation function as
\begin{align}
C_{\frac{1}{2}\frac{1}{2}\ell}(\De_1,\De_2,\De_3)
&:=
\lan \Psi_1(P_1,S_1) \bar\Psi_2(P_2,\bar S_2) O^{(\ell)}(P_3,Z_3) \ran
\nn\\
&=
f_{\frac{1}{2}\frac{1}{2}\ell}(P_i,\bar S_1,S_2,Z_3)
P_{12}^{-\frac{\Sig_{12}-\De_3+1-\ell}{2}}
P_{23}^{-\frac{\De_3-\De_{12}+\ell}{2}}
P_{31}^{-\frac{\De_3+\De_{12}+\ell}{2}}.
\label{1/21/2l}
\end{align}
The homogeneity conditions on the three fields \eqref{homoPS}, \eqref{homoT} 
and the transversality condition \eqref{transvTZ}
amount to the following constraints on $f_{\frac{1}{2}\frac{1}{2}\ell}$
\begin{align}
&{} 
f_{\frac{1}{2}\frac{1}{2}\ell}(\lm_iP_i, a\bar S_1, bS_2, cZ_3) 
= \lm_3^\ell abc^\ell \, f_{\frac{1}{2}\frac{1}{2}\ell}(P_i, \bar S_1, S_2, Z_3), 
\label{homo1/21/2l} \\
&{}
f_{\frac{1}{2}\frac{1}{2}\ell}(P_i, \bar S_1, S_2, Z_3) 
= f_{\frac{1}{2}\frac{1}{2}\ell}(P_i, \bar S_1, S_2, Z_3+\al P_3)
\quad \mbox{ for any } \, \al.
\label{transv1/21/2l}
\end{align}
Note that the homogeneity condition implies that $f_{\frac{1}{2}\frac{1}{2}\ell}$ 
must have only one spinor quadratics with $\bar S_1, S_2$.
We then find that solutions to the conditions \eqref{homo1/21/2l} and \eqref{transv1/21/2l} 
are linear combinations of the following three-point structures,
\begin{align}
f^{(1)}_{\frac{1}{2}\frac{1}{2}\ell} 
&:= 
\lan \bar S_1S_2\ran I_3^\ell, \\
f^{(2)}_{\frac{1}{2}\frac{1}{2}\ell}
&:= 
\lan \bar S_1P_3S_2\ran I_3^\ell \sqrt{\frac{P_{12}}{P_{13}P_{23}}}, \\
f^{(3)}_{\frac{1}{2}\frac{1}{2}\ell} 
&:= 
I_3^{\ell-1} J_3^{(1)}, \\
f^{(4)}_{\frac{1}{2}\frac{1}{2}\ell}
&:= 
I_3^{\ell-1} J_3^{(2)}, \\
f^{(5)}_{\frac{1}{2}\frac{1}{2}\ell} 
&:= I_3^{\ell-1} \lan \bar S_1Z_3P_3S_2\ran,
\end{align}
where we introduced several scalar quantities, 
all of which are defined to be invariant under the replacement $Z_3 \to Z_3+\al P_3$,
\begin{align}
I_3 &:= \frac{P_{23}(P_1.Z_3)-P_{13}(P_2.Z_3)}{P_{12}}, \\
J_3^{(1)} &:= \left[ \lan \bar S_1Z_3S_2\ran P_{13} + 2\lan \bar S_1P_3S_2\ran P_1.Z_3 \right]
\sqrt{\frac{P_{23}}{P_{12}P_{13}}}, \\
J_3^{(2)} &:= \left[ \lan \bar S_1Z_3S_2\ran P_{23} + 2\lan \bar S_1P_3S_2\ran P_2.Z_3 \right]
\sqrt{\frac{P_{13}}{P_{12}P_{23}}}.
\end{align}
For even-dimensional conformal field theories, 
the conditions \eqref{homo1/21/2l} and \eqref{transv1/21/2l} yield five more solutions,
which we denote by $f^{(i+5)}_{\frac{1}{2}\frac{1}{2}\ell}$ $(1 \leq i \leq 5)$, 
defined by replacing $\bar S_1$ in $f^{(i)}_{\frac{1}{2}\frac{1}{2}\ell}$ by $\bar S_1\Ga$.

So far, three-point functions with symmetric tensors have been considered,
while we can also compute those with tensors with antisymmetric indices.
Appendix \ref{app:antisymmetric} gives explicit three-point structures
in the three-point function of two spinor and one antisymmetric rank 2 tensor fields.

\paragraph{Differential representations}
It is known that three-point functions of two scalar and one tensor fields 
can be written as derivatives of scalar three-point functions \cite{Costa:2011dw}.
We can also find similar differential representations 
of the three-point function \eqref{1/21/2l}.\footnote{
Similar results for CFT$_3$ with Majorana spinors are found in \cite{Iliesiu:2015qra,Iliesiu:2015akf}
}
These expressions are useful when we compute conformal blocks with spinor fields, 
which will be explained briefly in Section \ref{block}.

For this, we first introduce symbols $C^{(i)}_{abc}$ 
for the three-point function \eqref{1/21/2l} 
with $f_{abc}$ replaced by $f^{(i)}_{abc}$.
Let us proceed to finding differential representations of them with respect to $P_1,P_2$.
The results in odd-dimensional cases are given by
\begin{align}
C^{(i)}_{\frac{1}{2}\frac{1}{2}\ell}(\De_1,\De_2,\De_3)
&= 
\bbD_{(i)}C_{00\ell}(\De_1,\De_2,\De_3)
\end{align}
where the differential operators $\bbD_{(i)}$ are defined by
\begin{align}
\bbD_{(1)} &:= \lan \bar S_1S_2 \ran \Pi_{\frac{1}{2},\frac{1}{2},0}, \\
\bbD_{(2)} &:= 
\frac{1}{2(\De_3-\ell-1)}\left[
\lan \bar S_1 \de_{P_1} S_2 \ran \Pi_{-\frac{1}{2},\frac{1}{2},0}
+ \lan \bar S_1 \de_{P_2} S_2 \ran \Pi_{\frac{1}{2},-\frac{1}{2},0}
\right], \\
\bbD_{(3)} &:= 
\frac{1}{2\ell}\left[
\lan \bar S_1 \de_{P_1} S_2 \ran \Pi_{-\frac{1}{2},\frac{1}{2},0}
- \lan \bar S_1 \de_{P_2} S_2 \ran \Pi_{\frac{1}{2},-\frac{1}{2},0}
\right] - \frac{\De_{12}-\ell}{\ell} \bbD_{(2)}, \\
\bbD_{(4)} &:= 
\frac{1}{2\ell}\left[
\lan \bar S_1 \de_{P_1} S_2 \ran \Pi_{-\frac{1}{2},\frac{1}{2},0}
- \lan \bar S_1 \de_{P_2} S_2 \ran \Pi_{\frac{1}{2},-\frac{1}{2},0}
\right] - \frac{\De_{12}+\ell}{\ell} \bbD_{(2)}, \\
\bbD_{(5)} &:= 
\frac{1}{2\ell(\De_3-1)} \Pi_{-\frac{1}{2},-\frac{1}{2},0},
\end{align}
where we introduced the differential operators $\de_{P_i}$ and the shift operators $\Pi_{a,b,c}$ by
\begin{align} 
&{} \de_{P_i}=\Ga^M\frac{\pd}{\pd P^M_i}, \\
&{} \Pi_{a,b,c}C_{00\ell}(\De_1,\De_2,\De_3)=C_{00\ell}(\De_1+a,\De_2+b,\De_3+c).
\end{align}
In even-dimensional cases, we need to add $C^{(i)}_{\frac{1}{2}\frac{1}{2}\ell}$ and $\bbD_{(i)}$
for $i=6,...,10$, which are obtained by replacing $\bar S_1$ by $\bar S_1\Ga$.
Here $C_{00\ell}$ is the three-point function of two scalar fields of dimension $\De_1,\De_2$ 
and one symmetric tensor field of dimension $\De_3$ and spin $\ell$, which is given by
\begin{align}
C_{00\ell}(\De_1,\De_2,\De_3) = I_3^\ell 
P_{12}^{-\frac{\Sig_{12}-\De_3-\ell}{2}}
P_{23}^{-\frac{\De_3-\De_{12}+\ell}{2}}
P_{31}^{-\frac{\De_3+\De_{12}+\ell}{2}}.
\end{align}
Therefore, the three-point function \eqref{1/21/2l} can be written 
in terms of scalar-scalar-tensor correlation functions,
\begin{align}
C_{\frac{1}{2}\frac{1}{2}\ell}(\De_1,\De_2,\De_3)
= \sum_{a} \lm^a_{\frac{1}{2}\frac{1}{2}\ell} \bbD_{(a)} C_{00\ell}(\De_1,\De_2,\De_3),
\label{diff-1/21/2l}
\end{align}
where the summation is taken over $a=1,...,5$ in odd dimensions
and over $a=1,...,10$ in even dimensions,
and $\lm^a_{\frac{1}{2}\frac{1}{2}\ell}$ are the unfixed coefficients for the three-point structures.

\subsection{Three-point functions with spinor, scalar and spinor-tensor primary fields}
Next, we consider a three-point function with spinor, scalar and spinor-tensor primary fields,
\begin{align}
\lan \Psi_1(P_1,S_1)\Phi(P_2)\bar O^{(\ell+\frac{1}{2})}(P_3,S_3,Z_3) \ran,
\end{align}
where the spinor field has dimension $\De_1$, the scalar field has dimension $\De_2$,
and the spinor-tensor field has dimension $\De_3$ and spin $\ell+\frac{1}{2}$.
Note that one spinor index and $\ell$ vector indices of $O^{(\ell+\frac{1}{2})}$ 
are contracted by $S_3$ and $Z_3$, respectively, to form a scalar polynomial in $Z_3$ of degree $\ell$.
The polarisation vector $Z_3$ is the same as the one for tensor fields. 

We first consider odd-dimensional conformal field theories as in the previous cases.
It is convenient to parametrise the correlation function as
\begin{align}
C_{\frac{1}{2},0,\ell+\frac{1}{2}}(\De_1,\De_2,\De_3)
&:=
\lan \Psi_1(P_1,S_1) \Phi(P_2) O^{(\ell+\frac{1}{2})}(P_3, S_3, Z_3) \ran
\nn\\
&=
f_{\frac{1}{2},0,\ell+\frac{1}{2}}(P_i,\bar S_1,S_3,Z_3)
P_{12}^{-\frac{\Sig_{12}-\De_3-\ell}{2}}
P_{23}^{-\frac{\De_3-\De_{12}+\ell}{2}}
P_{31}^{-\frac{\De_3+\De_{12}+\ell+1}{2}}.
\label{1/20l+1/2}
\end{align}
The homogeneity conditions on the three fields \eqref{homoPS}, \eqref{homoT}, \eqref{homoPS-st}
and the transversality condition \eqref{transvTZ}
amount to the following constraints on $f_{\frac{1}{2},0,\ell+\frac{1}{2}}$
\begin{align}
&{} 
f_{\frac{1}{2},0,\ell+\frac{1}{2}}(\lm_iP_i, a\bar S_1, bS_3, cZ_3) 
= \lm_3^\ell abc^\ell \, f_{\frac{1}{2},0,\ell+\frac{1}{2}}(P_i, \bar S_1, S_3, Z_3), 
\label{homo1/20l+1/2} \\
&{}
f_{\frac{1}{2},0,\ell+\frac{1}{2}}(P_i, \bar S_1, S_3, Z_3) 
= f_{\frac{1}{2},0,\ell+\frac{1}{2}}(P_i, \bar S_1, S_3, Z_3+\al P_3)
 \quad \mbox{ for any } \, \al.
\label{transv1/20l+1/2}
\end{align}
We then find that solutions to the conditions \eqref{homo1/20l+1/2} and \eqref{transv1/20l+1/2} 
are linear combinations of the following three-point structures,
\begin{align}
f^{(1)}_{\frac{1}{2},0,\ell+\frac{1}{2}} 
&:= \lan \bar S_1S_3\ran I_3^\ell, \\
f^{(2)}_{\frac{1}{2},0,\ell+\frac{1}{2}} 
&:= \lan \bar S_1P_2S_3\ran \sqrt{\frac{P_{31}}{P_{12}P_{23}}} I_3^\ell, \\
f^{(3)}_{\frac{1}{2},0,\ell+\frac{1}{2}} 
&:= \lan \bar S_1Z_3S_3\ran \sqrt{\frac{P_{13}P_{23}}{P_{12}}} I_3^{\ell-1}, \\
f^{(4)}_{\frac{1}{2},0,\ell+\frac{1}{2}} 
&:= \lan \bar S_1P_2Z_3S_3\ran \sqrt{\frac{P_{13}}{P_{12}}} I_3^{\ell-1}.
\end{align}
For even-dimensional conformal field theories, 
the conditions \eqref{homo1/21/2l} and \eqref{transv1/21/2l} yield four more solutions,
which we denote by $f^{(i+4)}_{\frac{1}{2},0,\ell+\frac{1}{2}}$ $(1 \leq i \leq 4)$, 
defined by replacing $\bar S_1$ in $f^{(i)}_{\frac{1}{2},0,\ell+\frac{1}{2}}$ by $\bar S_1\Ga$.

\paragraph{Differential representations}
As in the last subsection, we can find differential representations 
of the three-point correlation function \eqref{1/20l+1/2},
which are given by
\begin{align}
C^{(i)}_{\frac{1}{2},0,\ell+\frac{1}{2}}(\De_1,\De_2,\De_3)
&= 
\wt{\bbD}{}_{(i)}C_{00\ell}(\De_1,\De_2,\De_3),
\label{diff-1/20l+1/2}
\end{align}
where the differential operators $\bbD_{(i)}$ with respect to $P_1,P_2$ are defined as
\begin{align}
\wt{\bbD}{}_{(1)} &:= \lan \bar S_1S_3 \ran \Pi_{\frac{1}{2},0,\frac{1}{2}}, \\
\wt{\bbD}{}_{(2)} &:= 
\frac{1}{2(\Sig_{12}-\De_3+\ell-1)}\left[
\lan \bar S_1 \de_{P_1} S_3 \ran \Pi_{-\frac{1}{2},0,\frac{1}{2}}
+ \lan \bar S_1 \de_{P_2} S_3 \ran \Pi_{\frac{1}{2},-1,\frac{1}{2}}
\right], \\
\wt{\bbD}{}_{(3)} &:= 
-\frac{1}{\ell} \lan \bar S_1 \de_{P_2} S_2 \ran \Pi_{\frac{1}{2},-1,\frac{1}{2}}, \\
\wt{\bbD}{}_{(4)} &:= 
\frac{1}{\ell(\Sig_{12}-\De_3+\ell-2)} \lan \bar S_1 \de_{P_2}\de_{P_1} S_3 \ran 
\Pi_{-\frac{1}{2},-1,\frac{1}{2}}
- \frac{d+\ell+2+\De_{12}-\De_3}{\ell} \wt{\bbD}{}_{(1)}.
\end{align}
In even-dimensional cases, 
we need to add $C^{(i)}_{\frac{1}{2},0,\ell+\frac{1}{2}}$ and $\wt{\bbD}{}_{(i)}$ for $i=5,...,8$, 
which are obtained by replacing $\bar S_1$ by $\bar S_1\Ga$.
Therefore, the three-point function \eqref{1/20l+1/2} can be written 
in terms of scalar-scalar-tensor correlation functions,
\begin{align}
C_{\frac{1}{2},0,\ell+\frac{1}{2}}(\De_1,\De_2,\De_3)
= \sum_{a} \tilde{\lm}^a_{\frac{1}{2},0,\ell+\frac{1}{2}} \wt{\bbD}{}_{(a)} C_{00\ell}(\De_1,\De_2,\De_3),
\end{align}
where the summation is taken over $a=1,...,4$ in odd dimensions
and over $a=1,...,8$ in even dimensions,
and $\tilde{\lm}^a_{\frac{1}{2},0,\ell+\frac{1}{2}}$ are the unfixed coefficients for the three-point structures.

\subsection{Comments on conformal blocks with spinors}
\label{block}
We now proceed to the decomposition of four-point correlation functions of scalar and spinor fields 
in terms of conformal blocks. 
A four-point correlation function $\lan O_1O_2O_3O_4 \ran$ can be decomposed 
in terms of conformal blocks for primary fields $O$
appearing in the operator product expansions (OPE) $O_1 \times O_2$ and $O_3 \times O_4$,
\begin{align}
\lan O_1(P_1)O_2(P_2)O_3(P_3)O_4(P_4) \ran = 
\left( \frac{P_{24}}{P_{13}} \right)^{\frac{\De_{23}}{2}}
\left( \frac{P_{14}}{P_{13}} \right)^{\frac{\De_{34}}{2}}
\sum_{O;a,b} \frac{\lm_{12O} \lm_{34O} g^{a,b}_{\De,\ell}(u,v)}
{P_{12}^{\frac{\Sig_{12}+\ell_1+\ell_2}{2}} P_{34}^{\frac{\Sig_{34}+\ell_3+\ell_4}{2}}},
\end{align}
where $\lm^a_{12O}$ ($\lm^a_{34O}$) is the OPE coefficient of a primary field $O$ 
in the OPE $O_1 \times O_2$ ($O_3 \times O_4$),
the indices $a,b$ stand for the three-point structures,
the variables $u,v$ are the cross ratios,
and $(\De,\ell)$ are the conformal dimension and spin of $O$.

Among several methods to compute the conformal block $g^{a,b}_{\De,\ell}(u,v)$,
we adopt the method proposed in \cite{SimmonsDuffin:2012uy}. 
This method allows us to express conformal blocks as derivatives of scalar conformal blocks.
According to \cite{SimmonsDuffin:2012uy}, the conformal block for $O$ can be obtained 
by integrating the product of two three-point correlation functions, which reads
\begin{align}
&{}
\left( \frac{P_{24}}{P_{13}} \right)^{\frac{\De_{23}}{2}}
\left( \frac{P_{14}}{P_{13}} \right)^{\frac{\De_{34}}{2}}
\sum_{a,b}
\frac{\lm^a_{12O} \lm^b_{34O}}
{P_{12}^{\frac{\Sig_{12}+\ell_1+\ell_2}{2}} P_{34}^{\frac{\Sig_{34}+\ell_3+\ell_4}{2}}}
g^{ab}_{\De,\ell}(u,v)
\nn\\
&\quad\quad
= \frac{1}{N_O} \int \! D^dP \,
\lan O_1(P_1)O_2(P_2)O(P) \ran \lan \check{O}(P)O_3(P_3)O_4(P_4) \ran\big|_\cM,
\label{monodromy-int}
\end{align}
where $N_O$ is a normalisation factor,
$\check{O}$ is the shadow field of $O$, of dimension $d-\De$ and spin $\ell$,
and the symbol $|_\cM$ stands for the monodromy projection 
to filter out the correct conformal block from the integral.
In this paper, we consider the following two cases, 
one where the four fields are spinor fields of spin $1/2$, 
the other where $O_1,O_3$ are spinor fields while $O_2,O_4$ are scalar fields.

In the case where $O_1,O_2,O_3,O_4$ are spinor, 
the primary fields $O$ in the conformal block decomposition have integer spins,
while in the case where $O_1,O_3$ are spinor and $O_2,O_4$ are scalar, 
the primary fields $O$ have half-integer spins.
Since in the previous sections we have found 
the differential representations of the three-point correlation functions,
the integral \eqref{monodromy-int} can be rewritten 
as derivatives of that for scalar four-point correlation functions.
More explicitly, applying the differential representations \eqref{diff-1/21/2l}, \eqref{diff-1/20l+1/2} 
to the integral \eqref{monodromy-int}, we find that the conformal block $g^{a,b}_{\De,l}(u,v)$ satisfies
\begin{align}
&{}
\left( \frac{P_{24}}{P_{13}} \right)^{\frac{\De_{23}}{2}}
\left( \frac{P_{14}}{P_{13}} \right)^{\frac{\De_{34}}{2}}
\frac{1}{P_{12}^{\frac{\Sig_{12}+\ell_1+\ell_2}{2}} P_{34}^{\frac{\Sig_{34}+\ell_1+\ell_2}{2}}}
g^{a,b}_{\De,\ell}(u,v)
\nn\\
&\quad\quad
= \bbD^{12}_a \bbD^{34}_b \left[
\left( \frac{P_{24}}{P_{13}} \right)^{\frac{\De_{23}}{2}}
\left( \frac{P_{14}}{P_{13}} \right)^{\frac{\De_{34}}{2}}
\frac{1}{P_{12}^{\frac{\Sig_{12}}{2}} P_{34}^{\frac{\Sig_{34}}{2}}}
\bar{g}_{\De,\ell}(u,v)
\right],
\end{align}
where $\bar{g}_{\De,\ell}(u,v)$ is the conformal block with $(\De,\ell)$ 
for the scalar four-point correlation function of dimensions $\De_1,...,\De_4$.
The differential operators $\bbD^{12}_a$ 
are $\bbD_{(a)}$ for $(\ell_1,\ell_2)=(1/2,1/2)$,
and $\wt{\bbD}{}_{(a)}$ for $(\ell_1,\ell_2)=(1/2,0)$.
The differential operator $\bbD^{34}_a$ for each $a$ is obtained 
by applying the replacement $1 \to 3$ and $2 \to 4$ to $\bbD^{12}_a$,
noting that $\ell_1=\ell_3$ and $\ell_2=\ell_4$.

\section{Conclusion and Outlook}
\label{conclusion}
This paper computed three-point correlation functions with spinor, tensor, and spinor-tensor primary fields 
in general dimensional conformal field theories, without imposing any constraints on the spinors. 
In particular, we found all three-point structures of the correlation functions and their differential representations.
The differential representations were applied to compute conformal blocks 
associated with four-point functions with spinor and scalar fields.

As described in the Introduction, 
our final goal is to find geodesic Witten diagrams for the conformal blocks.
As shown in Section \ref{block}, the conformal blocks for four-point functions with spinors and scalars 
can be derived from scalar conformal blocks using the differential operators.
It is therefore straightforward to write the conformal blocks for four-point functions with scalars and spinors
as integrals over geodesics in AdS space 
since the geodesic Witten diagrams are known for scalar conformal blocks.
It, however, would be a nontrivial task to rewrite these integrals as geodesic Witten diagrams
with the bulk-boundary and bulk-bulk propagators for AdS spinor fields.
The first step in this direction would be 
to express all the three-point structures in Section \ref{correlators} as standard Witten diagrams
with the bulk-boundary propagators for AdS spinor fields, 
which have been done for three-point functions of tensor fields of arbitrary integer spins 
in \cite{Castro:2017hpx,Dyer:2017zef,Sleight:2017fpc,Chen:2017yia},
using AdS propagators provided in  \cite{Costa:2011dw}.
This enables us to pin down bulk interaction vertices for the three-point structures.
To perform this analysis, we would need to rewrite the AdS$_{d+1}$ propagators
in the $(d+2)$-dimensional Dirac matrices for the embedding space. 
It amounts to extending the embedding space formalism 
to cover AdS spinor fields.
We hope to report our progress in this direction in the near future.

\section*{Acknowledgements}
This work is supported by the ``CUniverse'' research promotion project by Chulalongkorn University (grant reference CUAASC). The author is grateful to Auttakit Chatrabhuti, Andrea Guerrieri, Rongvoram Nivesvivat, Toshifumi Noumi, Tadashi Okazaki, and Yuki Sato for fruitful discussions and suggestions.

\appendix
\section{Tensor primary fields of integer spins}
\label{app:pol-tensor}
We give a brief review of the embedding of a tensor primary field $t_{\mu_1...\mu_\ell}(x)$ of integer spin $\ell$ 
into the embedding space.
Here we just consider symmetric and traceless tensors.
For this, we introduce a symmetric tensor field $T_{M_1...M_\ell}(P)$ in the embedding space.
Its values at $P$ and at $\lm P$, which are geometrically identical, are related by the homogeneity condition
\begin{align}
\label{homoT}
T_{M_1...M_\ell}(\lm P) = \lm^{-\De} T_{M_1...M_\ell}(P).
\end{align}
For the tensor to be tangent to the null hypersurface, we need another condition, 
which is the transversality condition
\begin{align}
\label{transvT}
P^{M_k}T_{M_1...M_\ell}(P)=0 \quad\quad \mbox{ for } \quad 1 \leq k \leq \ell.
\end{align}
To compute correlation functions, it is convenient to introduce the polarization vector $Z^M$,
satisfying $Z^2=Z.P=0$,
to form scalar fields.\footnote{
Tensors of rank $\ell$ proportional to vectors $P_{M_i}$ ($1 \leq i \leq \ell$) 
do not survive the projection to the physical space.
This implies that two tensors differing by tensors proportional to $P_{M_i}$ 
can be considered to be ``gauge-equivalent'' 
because the two become an identical tensor field in the physical space.
The condition $Z.P=0$ is imposed to guarantee the gauge invariance of scalar polynomial in $Z$, 
meaning that two gauge equivalent tensors are mapped to an identical scalar polynomial in $Z$.
}
For example, a tensor field $T_{M_1...M_\ell}(P)$ can be encoded in a polynomial in $Z$ of degree $\ell$,
\begin{align}
T(P, Z) = Z^{M_1} Z^{M_2} ... Z^{M_\ell} T_{M_1M_2 ... M_\ell}(P).
\end{align}
Note that this polynomial is Lorentz scalar, which makes computations of correlation functions much simpler.
In terms of this polynomial, the transversality condition \eqref{transvT} becomes
\begin{align}
\label{transvTZ}
T(P, Z) = T(P, Z+\al P) \quad \mbox{ for any } \, \al.
\end{align}
We can recover the original tensor field in the embedding space by 
\begin{align}
\label{recover-tensor}
T_{M_1...M_\ell}(P) = \cD_{M_1}...\cD_{M_\ell} T(P, Z),
\end{align}
where the derivative operator $\cD_M$ is defined as
\begin{align}
\cD_M := 
\left( \frac{d-2}{2}+Z^N\frac{\pd}{\pd Z^N} \right) \frac{\pd}{\pd Z^M}
- \frac{1}{2}Z_M \frac{\pd^2}{\pd Z^N \pd Z_N}.
\label{recover-tensor-diffop}
\end{align}
We can then recover the original tensor field in the physical space
by replacing coordinates $P^M$ by $P^M(x)$ 
and multiplying each (lowered) vector index $M$ of the embedding-space tensor field 
by $\pd P^M(x)/\pd x^\mu$.

The polarisation vector simplifies the computation of correlation functions of tensor fields
because scalar correlation functions are much easier to compute.
The dependence on polarisation vectors are fixed by the homogeneity and transversality conditions.
A two-point function of two symmetric tensor fields of dimension $\De$ and spin $\ell$ is given by
\begin{align}
\lan O^{(l)}(P_1,Z_1)O^{(\ell)}(P_2,Z_2) \ran
= \left[
(Z_1.Z_2)P_{12} + 2(Z_1.P_2)(Z_2.P_1)
\right]^\ell P_{12}^{-\De-\ell}
\end{align}
A three-point correlation function of two scalar fields of dimensions $\De_1,\De_2$ 
and one symmetric tensor field of dimension $\De_3$ and spin $\ell$ is given by
\begin{align}
\lan \Phi_1(P_1) \Phi_2(P_2) O^{(\ell)}(P_3,Z_3) \ran = I_3^\ell 
P_{12}^{-\frac{\Sig_{12}-\De_3-\ell}{2}}
P_{23}^{-\frac{\De_3-\De_{12}+\ell}{2}}
P_{31}^{-\frac{\De_3+\De_{12}+\ell}{2}}.
\end{align}
For more details, see, for example, \cite{Costa:2011mg}.

\section{Embeddings of the Dirac matrices}
\label{app:embed}
We here give explicit forms of the embeddings of the $d$-dimensional Dirac matrices 
into the $(d+2)$-dimensional ones. 
As described in the main part, we only consider the two typical cases given below.  
\paragraph{Euclidean metric} 
We define the $(d+2)$-dimensional Dirac matrices $\{\Ga^M\}$ as
\begin{align}
\Ga^\mu = \begin{pmatrix} \ga^\mu & 0 \\ 0 & -\ga^\mu \end{pmatrix}, \quad
\Ga^\sharp = \begin{pmatrix} 0 & 1 \\ 1 & 0 \end{pmatrix}, \quad
\Ga^\flat = \begin{pmatrix} 0 & 1 \\ -1 & 0 \end{pmatrix},
\end{align}
Note that $\Ga^\flat$ is anti-Hermitian and the rest, including $\ga^\mu$, are Hermitian.
We then define the $(d+2)$-dimensional Dirac conjugate matrix $D$ by
\begin{align}
D = \begin{pmatrix} 0 & -1 \\ 1 & 0 \end{pmatrix},
\end{align}
which satisfies
\begin{align}
D^\dg = D^{-1} = -D, \quad D \Ga^M D^{-1} = -(\Ga^M)^\dg.
\end{align}
Note that $D$ is different from $D$ for Lorentzian metric by factor $-1$.

\paragraph{Minkowski metric with one minus sign}
We define $\{\Ga^M\}$ as
\begin{align}
\Ga^\mu = \begin{pmatrix} \ga^\mu & 0 \\ 0 & -\de \ga^\mu \de^{-1} \end{pmatrix}, \quad
\Ga^\sharp = \begin{pmatrix} 0 & \de^{-1} \\ \de & 0 \end{pmatrix}, \quad
\Ga^\flat = \begin{pmatrix} 0 & \de^{-1} \\ -\de & 0 \end{pmatrix},
\end{align}
where we introduced the $d$-dimensional Dirac conjugate matrix $\de:=\ga^0$, 
which satisfies
\begin{align}
\de^\dg = \de^{-1} = -\de, \quad \de\ga^\mu\de^{-1} = -(\ga^\mu)^\dg.
\end{align}
Note that $\ga^0, \Ga^0,\Ga^\flat$ are taken to be anti-Hermitian and the rest are Hermitian.
We then introduce the $(d+2)$ Dirac conjugate matrix $D$ by
\begin{align}
D = \begin{pmatrix} 0 & 1 \\ -1 & 0 \end{pmatrix},
\end{align}
which satisfies
\begin{align}
D^\dg = D^{-1} = -D, \quad D \Ga^M D^{-1} = (\Ga^M)^\dg.
\end{align}
Here note that the Dirac conjugation matrix $D$ in the two cases have opposite signs
so that all results in the Euclidean case are obtained just by setting $\de=1$
in the corresponding results in the Lorentzian case.
This is possible by the choice of $D$ such that 
$D$ in the Euclidean case has the opposite sign to that in the Lorentzian case.
 
\section{Three-point functions with a tensor field with antisymmetric indices}
\label{app:antisymmetric}
In the case with a totally symmetric tensor, 
spinor quadratics with antisymmetrised products of Dirac matrices were excluded.
If the tensor field is anti-symmetric for some indices, then we need such spinor quadratics.
To see it explicitly, let us consider the case where $O$ is a totally antisymmetric rank 2 tensor.
Here we do not use the polarisation vectors.
The homogeneity and transversality conditions are the same as the totally symmetric cases.
We then find that solutions to the constraints 
are linear combinations of the following three-point structures\footnote{
We introduced $I_M, J^{(1)}_M, J^{(2)}_M$ by
\begin{align}
I_3=Z_3^MI_M, \quad
J^{(1)}_3=Z_3^MJ^{(1)}_M, \quad
J^{(2)}_3=Z_3^MJ^{(2)}_M, \quad \nn
\end{align}
}:
\begin{align}
\big( f^{(1a)}_{\frac{1}{2}\frac{1}{2}2} \big)_{MN} 
&:= I_MJ^{(1)}_N-I_NJ^{(1)}_M, \\
\big( f^{(2a)}_{\frac{1}{2}\frac{1}{2}2} \big)_{MN} 
&:= I_MJ^{(2)}_N-I_NJ^{(2)}_M, \\
\big( f^{(3a)}_{\frac{1}{2}\frac{1}{2}2} \big)_{MN} 
&:= I_M\lan \bar S_1P_3\Ga_NS_2\ran-I_N\lan \bar S_1P_3\Ga_MS_2\ran, \\
\big( f^{(4a)}_{\frac{1}{2}\frac{1}{2}2} \big)_{MN} 
&:= 
\lan \bar S_1 \Ga_{MN} S_2 \ran \frac{P_{13}P_{23}}{P_{12}}
+ \lan \bar S_1 [\Ga_M(P_1)_N-\Ga_N(P_1)_M,P_3] S_2 \ran \frac{P_{23}}{P_{12}}, \\
\big( f^{(5a)}_{\frac{1}{2}\frac{1}{2}2} \big)_{MN} 
&:= 
\lan \bar S_1 \Ga_{MN} S_2 \ran \frac{P_{13}P_{23}}{P_{12}}
+ \lan \bar S_1 [\Ga_M(P_2)_N-\Ga_N(P_2)_M, P_3] S_2 \ran \frac{P_{13}}{P_{12}}, \\
\big( f^{(6a)}_{\frac{1}{2}\frac{1}{2}2} \big)_{MN} 
&:= \lan \bar S_1 \{\Ga_{MN}, P_3\} S_2 \ran,
\end{align}
where $\Ga_{MN}=\frac{1}{2}[\Ga_M,\Ga_N]$.
For even-dimensional conformal field theories, 
the homogeneity and transversality conditions yield six more solutions, 
which are obtained by replacing $\bar S_1$ in each expression by $\bar S_1\Ga$.

\end{document}